\documentstyle[12pt,a4wide]{article}
\normalsize
\def\simless{\mathbin{\lower 3pt\hbox{$\rlap{\raise 5pt\hbox{$\char'074$}}
\mathchar"7218$}}}
\def\simgreat{\mathbin{\lower 3pt\hbox{$\rlap{\raise 5pt \hbox{$\char'076$}}
\mathchar"7218$}}}
\catcode`@=11

\def\beqra{\begin{eqnarray}} \def\eeqra{\end{eqnarray}}
\def\beq{\begin{equation}}      \def\eeq{\end{equation}}
\def\ds{\displaystyle}
\def\ts{\textstyle}

\def\fo{\hbox{{1}\kern-.25em\hbox{l}}}

\def\ch{\@startsection{section}{1}{\z@}{-3ex plus-1ex minus-.2ex}%
        {2ex plus.2ex}{\large\sc}}

\def\; \lapp \;{\raisebox{-.4ex}{\rlap{$\sim$}} \raisebox{.4ex}{$<$}}

\def\con{\ifmmode \hbox{\bf*} \else{\bf*}\fi}   
\def\scon{\ifmmode \hbox{\footnotesize\rm\bf*} \else{\footnotesize\rm\bf*}\fi}

\def\0#1{\relax\ifmmode\mathaccent"7017{#1}
        \else\accent23#1\relax\fi}              

\def\eslash{\not{\hbox{\kern-2pt $E$}}}

\catcode`@=12
\input feynman
\begin{document}
\hoffset=0.4cm
\voffset=-1truecm
\normalsize


\begin{titlepage}
\begin{flushright}
DFPD 94/TH/39\\
SISSA 94/82-A\\
\end{flushright}
\vspace{24pt}
\centerline{\Large {\bf On Spontaneous Baryogenesis and Transport Phenomena}}
\vspace{24pt}
\begin{center}
{\large\bf D. Comelli$^{a,b,}$\footnote{Email:
comelli@mvxtst.ts.infn.it; supported by Schweizerischen national
funds.}
, M. Pietroni$^{c,}$\footnote{Email: pietroni@mvxpd5.pd.infn.it; address after
September 1, 1994: Deutsches Elektronen - Synchrotron, DESY, Hamburg, Germany}
 and A. Riotto$^{c,d,}$\footnote{Email: riotto@tsmi19.sissa.it; on leave of
absence from International School for Advanced Studies (ISAS), Trieste, Italy}}
\end{center}
\vskip 0.2 cm
{\footnotesize
\centerline{\it $^{(a)}$Dipartimento di Fisica Teorica Universit\`a di
Trieste,}
\centerline{\it Strada Costiera 11, 34014 Miramare, Trieste, Italy}
\vskip 0.2 cm
\centerline{\it $^{(b)}$ Istituto Nazionale di Fisica Nucleare,}
\centerline{\it Sezione di Trieste, 34014 Trieste, Italy}
\vskip 0.2 cm
\centerline{\it $^{(c)}$Istituto Nazionale di Fisica Nucleare,}
\centerline{\it Sezione di Padova, 35100 Padua, Italy.}
\vskip 0.2 cm
\centerline{\it $^{(d)}$Instituto de Estructura de la Materia, CSIC,}
\centerline{\it Serrano, 123, E-20006 Madrid, Spain}}
\vskip 0.5 cm
\centerline{\large\bf Abstract}
\vskip 0.2 cm
\baselineskip=15pt
The spontaneous baryogenesis mechanism by Cohen, Kaplan and Nelson is
reconsidered taking into account the transport of particles inside
the electroweak bubble walls. Using linear response theory,  we calculate
the modifications
on the thermal averages of the charges of the system due to the
presence of a space time dependent `charge potential' for a quantum
number not orthogonal to baryon number.
The local equilibrium configuration is discussed,
showing that, as a consequence of  non zero densities for conserved
charges, the $B+L$ density is driven to non zero values by sphaleronic
processes.

Solving a rate equation for the baryon number generation, we
obtain an expression for the final baryon asymmetry of the Universe
containing the relevant parameters of the bubble wall, {\it i.e.} its
velocity, its width, and the width of the region in which sphalerons are
active. Compared to previous estimates in which transport effects were not
taken into account, we find an enhancement of nearly three orders of magnitude
in the baryon asymmetry.

Finally, the role of QCD
sphalerons in cooperation with transport effects is analyzed, showing that
the net result depends crucially on the particle species which enter into
the  charge potential.

\end{titlepage}

\baselineskip=18pt
\setcounter{page}{1}
\begin{center}
{{\Large {\bf 1. Introduction}}}
\end{center}
\vspace{1. truecm}
The possibility of a baryogenesis at the electroweak scale is a very popular
but
controversial topic. Despite the large number of related publications
\cite{reviews}, none of the key aspects of this subject can be considered to
be on firm ground. First, it is well known that the requirement that the
anomalous `sphaleronic' processes which violate baryon number ($B$) go out of
equilibrium soon after the transition translates into a lower bound
on the
ratio between the vacuum expectation value (VEV) of the Higgs field, $v(T)$,
and
the critical temperature of the transition, $T_c$, $v(T_c)/T_c \simgreat 1$.
In the
standard model, improved perturbative evaluations of this ratio \cite{smpert}
give a value
which is badly less than unity for values of the mass of the Higgs scalar
compatible with LEP results. The situation in the minimal supersymmetric
standard model is slightly better, however the allowed region of the parameter
space is  very small, and is likely to be excluded by future LEP data
\cite{MSSM}. On the other hand, the perturbative expansion cannot be trusted
any more for values of the Higgs masses of the order of the $W$ boson
mass or larger \cite{pertfail}, and preliminary results based on
non perturbative methods (lattice simulations  \cite{lattice},
$1/\varepsilon$ expansion
\cite{epsilon}, effective action \cite{alford}) give indications of
strong differences of the results with respect to those obtained
perturbatively.
Clearly, much work is still needed on this
issue.

Another aspect of the problem is CP violation. This has been the subject of
a recent debate in the literature
about the need of further complex phases in the theory besides the one in the
Cabibbo-Kobayashi-Maskawa mixing matrix (see. \cite{CPSM}). In  models
with more than one Higgs doublet, like the minimal supersymmetric standard
model, further sources of CP violation can emerge naturally from the Higgs
sector. In particular, the possibility of a spontaneous CP violation at
finite temperature has been emphasized \cite{scpv, quiros, singlet}.
This effect could
give enough contribution for the baryogenesis  and at the same time satisfy
the upper bounds on CP violation coming from the electric dipole moment
of the neutron.

Even assuming that the phase transition is strong and that CP
violation  is enough, we must face the other key issue, namely what is the
mechanism responsible for the generation of baryons during the phase
transition.
The most significant departure from  thermodynamic equilibrium takes place
at the passage of the walls of the expanding bubbles which convert
the unbroken into the broken phase. According to the size and speed of the
bubble walls, two different mechanisms are thought to be dominant. In the case
of ``thin" (width $\sim 1/T$) walls, typical of a very strong phase transition,
the creation of baryons occurs via the asymmetric (in baryon number)
reflection of quarks from the bubble wall, which biases the sphaleronic
transitions in the region in front of the expanding bubble \cite{reflection}.
If the walls are ``thick" (width$\sim (10 - 100)/T$) then the relevant
mechanism takes place inside the bubble walls rather than in front of them. In
this
case we can make a distinction between `fast' processes (mediated by gauge,
flavour diagonal, interactions and by top Yukawa interactions) and
`slow' processes
(mediated by Cabibbo suppressed gauge interactions and by light quarks
Yukawa interactions). The
former are able to follow adiabatically the changing of the Higgs VEV inside
the bubble wall, while, in first approximation, the latter are frozen during
the passage of the wall. If CP violation, explicit or spontaneous, is present
in
the scalar sector then a space-time dependent phase for the Higgs VEVs is
turned on inside the wall. The time derivative of this phase couples with
the density of a quantum number non orthogonal to baryon number (e.g. fermion
hypercharge density) \footnote{The phase also couples to Chern-Simons
number but this coupling induces an effect which is suppressed by
 $m_{top}(T)^2/T^2$ with respect to the one which we are presently discussing
\cite{chernsimons}, so we will neglect it.} and then can be seen as an
effective chemical potential, named ``charge potential", which has the effect
of biasing the rates of the sphaleronic processes, creating an asymmetry
proportional to $\dot{\vartheta}$, where $\vartheta$ is the phase of the VEVs.

This ``adiabatic scenario", originally due to Cohen, Kaplan, and Nelson
(CKN) \cite{CKN}, has been
recently reconsidered by different authors in different but related aspects.
First, Giudice and Shaposhnikov have shown the dramatic effect of non
perturbative, chirality breaking, transitions induced by the so called
``QCD sphalerons"  \cite{GianShap}. If these processes were active inside
the bubble walls, then
the equilibrium value for baryon number in the adiabatic approximation would be
proportional to that for the conserved quantum  number $B-L$ ($L$ is
the lepton number), up to mass effects suppressed by
$\sim (m_{top}(T)/\pi T)^2$.
Then, imposing the constraint $\langle B-L \rangle = 0$ (here $\langle \cdots
\rangle$ represents the thermal average) we obtain zero baryon number (up to
mass effects).

Dine and Thomas \cite{dinethomas} have considered the two Higgs doublets
model in
which the same doublet couples both to up and down quarks, the same model
considered in the original work by CKN. These authors have pointed out that
$\dot{\vartheta}$ couples also to the Higgs density, so that the induced
charge potential is for total hypercharge rather than for fermion hypercharge.
As long as effects proportional to the temperature dependent VEV $v(T)$ are
neglected, hypercharge is a exactly conserved quantum number and then, again
imposing the constraint that all the conserved charges have zero thermal
averages, no baryon asymmetry can be generated. So, we again  find a
$m_{top}(T)^2/T^2$ suppression factor.

Finally, Joyce, Prokopec and Turok (JPT) have emphasized the very
important point
that the response of the plasma to the charge potential induced by
$\dot{\vartheta}$
is not simply that of a system of fixed charges, because transport phenomena
may play a crucial role \cite{JPT}. When a space time dependent charge
potential
is turned on at a certain point, hypercharged particles are displaced from
the surrounding regions, so that even the thermal averages of conserved quantum
numbers become locally non vanishing. As a consequence, the equilibrium
properties of the system have to be reconsidered taking into account the local
`violation' of the conserved quantum numbers.

In this paper we analyze the adiabatic scenario using linear response
theory \cite{LRT, Kapusta} in order to take  transport effects into account.
We assume that a spacetime dependent charge potential for fermion
hypercharge is generated inside the bubble wall, without discussing
its origin, and investigate its effect on the thermal averages of the
various quantum numbers of the system.
We find that transport phenomena are really
crucial, but we disagree with JPT's  conclusion that as a consequence
of the local `violation' of global quantum numbers there is no biasing of
the sphaleronic processes. Actually, in the adiabatic approximation the
local equilibrium configuration of the system is determined by the thermal
averages of the charges conserved by all the  fast interactions. The effect of
transport phenomena is to induce  space time dependent non zero values for
these averages. We calculate these averages using linear response theory
 and then determine the local equilibrium configuration,
showing
that it corresponds to $\langle B+L\rangle \neq 0$. As we will
discuss, JPT's result corresponds to freezing out any interaction
inside the bubble wall, which is in contradiction with the adiabatic
hypothesis. Then we write down a rate
equation in order to take into account the slowness of the sphaleron
transitions
and obtain an expression for the final baryon asymmetry explicitly containing
the parameters describing the bubble wall, such as its velocity, $v_w$, its
width, and the width of the region in which the sphalerons are active.

The inclusion of transport phenomena also sheds a new light on the strong
sphaleron effects and on the effect of a charge potential for total
rather than fermionic hypercharge. The dramatic suppressions
found by Giudice and Shaposhnikov and by Dine and Thomas respectively,
are both a consequence of taking zero averages for
conserved quantum numbers. Since these averages are no more locally zero we
will find a non zero  $\langle B+L\rangle \neq 0$, even in the case in
which the charge potential is for total
rather than for fermion hypercharge.
In the case of QCD sphalerons we find that the final result depends in a
crucial
way on the form of the charge potential which is considered. For example, if
all the left handed fermions plus the right handed quarks contributed to the
charge potential according to their hypercharge, then no bias of sphaleron
processes would be obtained. In this case, we would find a  non zero value
for B+L inside the bubble wall but no baryon asymmetry far from it inside the
broken phase. On the other hand, if only right and left handed top quarks
participate to the charge potential, then a final asymmetry is found and
QCD sphalerons are harmless.

 The plan of the paper is as follows: in section 2 we will develop a chemical
potential analysis for the equilibrium properties of the plasma inside
the bubble wall in the adiabatic approximation, and we will write
down the rate equation fro the production of $B+L$ due to sphaleron
transitions;
in sect. 3 we will introduce our application of linear response analysis to the
calculation of the variations of the thermal averages induced by
$\dot{\vartheta}$.
In particular we will show that the fundamental quantity to evaluate is
the retarded two points Green's function for fermion currents. In section 4
we will solve the rate equation, finding an expression for the final baryon
asymmetry in terms of the various bubble wall parameters.
In this context, we will also discuss the screening effects on the electric
charge. In sect. 5 we  will discuss the role of QCD sphalerons in
cooperation with transport phenomena, and finally we will discuss our
results in section 6.

\vspace{.5 cm}
{\Large\center\bf{2. Local equilibrium inside the wall}}
\vspace{1. truecm}

For definiteness, we will work in the two Higgs doublet model in which one
doublet couples to up quarks and the other one to down quarks. The phase
transition is assumed to be strong enough so that sphaleron processes freeze
out
somewhere inside the bubble wall (see the end of sect. 4 and ref.
\cite{dinethomas} for a discussion about this point).

The relevant timescale for baryogenesis is given by the passage of the bubble
wall, which takes $\Delta t_w = \Delta z /v_{w} \simeq (200)/T$ \cite{thick},
where $\Delta z$ is the wall thickness. During this time the phase of the Higgs
VEV's changes of an amount $\Delta \vartheta$. Thus, we can discriminate
between
fast processes, which have a rate $\simgreat 1/ \Delta t_w$ and then can
equilibrate adiabatically with $\dot{\vartheta}$, and slow interactions, which
feel that $\vartheta$ is changing only when the bubble has already passed
by and
sphalerons are no more active. Next, we introduce a chemical potential for any
particle which takes part to fast processes, and then reduce the number
of linearly independent chemical potentials by solving the corresponding system
of equations, in a way completely analogous to that followed for example in
refs. \cite{HarveyDreiner}, the main difference here being that light
quark Yukawa interactions and Cabibbo suppressed gauge interactions
are out of equilibrium.  Finally, we can express the abundances of any
particle in equilibrium in terms of the remaining linear independent chemical
potentials, corresponding to the conserved charges of the system.

Since strong interactions are in equilibrium inside the bubble wall, and since
the current coupled  to $\dot{\vartheta}$ is color singlet, we  can chose the
same chemical potential for quarks of the same flavour but different color, and
set to zero the chemical potential for gluons.
Moreover, since inside the bubble wall $SU(2)\times U(1)$ is broken, the
chemical potential for the neutral Higgs scalars vanishes\footnote{This is
true if chirality flip interactions, or processes like $Z\rightarrow Z^* h$,
are
sufficiently fast; since the corresponding rates depend on the Higgs VEV,
they will be suppressed by factors of $(v(T)/T)^2$ with respect for example to
 the rate for $h t_L \leftrightarrow t_R g$. This has led the authors of
refs.\cite{GianShap, JPT} to consider the system  in the unbroken phase.
Anyway,
this choice does not lead to dramatic changes to the conclusion of this
paper.}.

The other fast processes, and the corresponding chemical potential equations
are:\\
i) top Yukawa:
\beq
\begin{array}{ccrl}
t_L + H_2^0 \leftrightarrow t_R + g &
\:\:\:\:\:&
(\mu_{t_L} =& \mu_{t_r})\\
b_L + H^+ \leftrightarrow t_R + g&
\:\:\:\:\:&
(\mu_{t_R}=&\mu_{b_L} + \mu_{H^+})
\end{array}\\
\label{chem1}
\eeq
ii) SU(2) flavour diagonal:
\beq
\begin{array}{ccrl}
e_L^i\leftrightarrow \nu_L^i + W^-&
\:\:\:\:\:&
(\mu_{\nu_L^i} =& \mu_{e_L^i} + \mu_{W^+})\\
u_L^i\leftrightarrow d_L^i + W^+&
\:\:\:\:\:&
(\mu_{u_L^i} =& \mu_{d_L^i} + \mu_{W^+})\\
H_2^0 \leftrightarrow H^+ + W^-&
\:\:\:\:\:&
(\mu_{H^+} =&\mu_{W^+})\\
H_1^0 \leftrightarrow H^- + W^+&
\:\:\:\:\:&
(\mu_{H^-} =&-\mu_{W^+})\\
\end{array}
(i=1,\:2,\:3)
\label{chem2}
\eeq
Neutral current gauge interactions are also in equilibrium, so we have zero
chemical potential for the photon and the $Z$ boson.

Imposing the above constraints, we can reduce the number of
independent chemical potentials to four, $\mu_{W^+}$, $\mu_{t_L}$,
$\mu_{u_L}\equiv 1/2 \sum_{i=1}^2 \mu_{u_L^i}$, and $\mu_{e_L}\equiv
1/3 \sum_{i=1}^3 \mu_{e_L^i}$. These quantities correspond to the four
linearly independent conserved charges of the system. Choosing the
basis $Q'$, $(B-L)'$, $(B+L)'$, and $BP'\equiv B_3' - 1/2
(B_1'+B_2')$, where the primes indicate that only particles in
equilibrium contribute to the various charges, and introducing the
respective chemical potentials, we can go to the new basis using the
relations
\beq
\left\{
\begin{array}{ccl}
\mu_{Q'} & =& 3 \mu_{t_L} + 2 \mu_{u_L} - 3 \mu_{e_L} + 11 \mu_{W^+}\\
\mu_{(B-L)'} &=& 3 \mu_{t_L} + 4 \mu_{u_L} - 6 \mu_{e_L} - 6 \mu_{W^+}\\
\mu_{(B+L)'} &=& 3 \mu_{t_L} + 4 \mu_{u_L} + 6 \mu_{e_L}\\
\mu_{BP'} &  =& 3 \mu_{t_L} - 2 \mu_{u_L}
\end{array}
\right.
\label{conserva}
\eeq

If sphaleron transitions were fast, then we could eliminate a further
chemical potential through the constraint
\beq
2 \sum_{i=1}^3 \mu_{u_L^i} +  \sum_{i=1}^3 \mu_{d_L^i} +  \sum_{i=1}^3
\mu_{e_L^i}
= 0.
\label{sfalc}
\eeq
In this case, the value of $(B+L)'$ would be determined by that of the
other three charges according to the relation
\beq
(B+L)_{EQ}' = \frac{3}{80} Q' + \frac{7}{20} BP' - \frac{19}{40}
(B-L)'
\label{equilibrium}
\eeq
where  we have indicated charge densities by the corresponding charge
symbols.  We have
neglected mass effects, which means that the excess of particle
over antiparticle density is related to the chemical potentials according
to the relation \cite{HarveyDreiner} $n_+ - n_- =  a g T^3/6 (\mu/T)$,
where $a=1$
for fermions and $a=2$ for bosons, and $g$ is the number of spin and
color degrees of freedom.

The above result should not come as a surprise, since we already know
from ref. \cite{HarveyDreiner} that a non zero value for $B-L$ gives
rise to a non zero $B+L$ at equilibrium. Stated in other words,
sphaleron transitions
erase the baryon asymmetry only if any conserved charge of the system
has vanishing thermal average, otherwise the equilibrium point
lies at $(B+L)_{EQ} \neq 0$.

Actually, sphaleron rates are too small to allow $(B+L)'$ to reach its
equilibrium value (\ref{equilibrium}), $\tau_{sp} \simeq$
$(\alpha_W^4 T)^{-1} \gg \Delta t_W$, so equilibrium thermodynamics
cannot be used to describe baryon number generation inside the bubble
wall. Following refs.  \cite{shap, CKN} we shall make use of the rate
equation
\beq
\frac{d\:}{d t} (B+L)_{SP}' = - \frac{\Gamma_{SP}}{T} \frac{\partial
F}{\partial (B+L)'}
\label{rate}
\eeq
where $\Gamma_{SP}=k (\alpha_W T)^4 \exp(-\phi/g_W T)$ is the rate of
the sphaleron transitions when the value of the Higgs field is $\phi$
($k\simeq (0.1 - 1)$ from numerical simulations \cite{k}), $F$ is the free
energy
of the system, and the derivative with respect to $(B+L)'$ must be
taken keeping $Q'$, $(B-L)'$, and $BP'$, constant. The meaning of eq.
(\ref{rate}) is straightforward. Sphaleron transitions (which change
$(B+L)'$ but conserve $Q'$, $(B-L)'$, and $BP'$) will be turned on only
if they  allow the {\it total} free energy of the system to get closer to its
minimum, {\it i.e.} equilibrium, value.
At high temperature $(\mu_i \ll T)$ the free energy of the system is
given by
\beq
\begin{array}{rl}
F = \frac{T^2}{12} &\left[3 \mu_{e_L}^2 + 3 \mu_{\nu_L}^2 +
6\mu_{u_L}^2 + 3 \mu_{t_L}^2 + 3\mu_{t_R}^2 + 6\mu_{d_L}^2 +
3\mu_{b_L}^2 \right.\\
&\left.+ 6\mu_{W^+}^2 + 2\mu_{H^+}^2 + 2\mu_{H_1^0}^2 +
2\mu_{H_2^0}^2\right].
\end{array}
\eeq
Using (\ref{chem1}), (\ref{chem2}) and (\ref{conserva}) to express
the chemical potentials in terms of the four conserved charges in
(\ref{conserva}) we obtain the free energy as a function  of the
density of $(B+L)'= \mu_{B+L} T^2/6$,
\beq
F\left[(B+L)'\right] = 0.46 \frac{\left[(B+L)' -
(B+L)'_{EQ}\right]^2}{T^2} + constant \:\:terms
\eeq
where the {\it ``constant terms"} depend on $Q'$, $(B-L)'$, and $BP'$
but not on $(B+L)'$,
and $(B+L)'_{EQ}$ is given by (\ref{equilibrium}). The total amount of
$(B+L)'$ present in a certain point at a certain time is made up by
two contributions: $(B+L)'_{SP}$, generated by sphaleron transitions, and
$(B+L)'_{TR}$, which is not generated but is transported from nearby regions in
response to the perturbation introduced by $\dot{\vartheta}$.
So, eq. (\ref{rate}) now takes the form
\beq
\frac{d\:}{d t} {(B+L)'_{SP}} = - 0.92 \frac{\Gamma_{SP}}{T^3}
\left[ (B+L)'_{SP} +
(B+L)'_{TR} - (B+L)'_{EQ} \right]
\label{rate1}
\eeq
with the initial condition $(B+L)'_{SP}=0$ before $\dot{\vartheta}$ is
turned on.

Let us summarize our discussion up to this point. Consider an
observer in the plasma reference frame during the phase transition.
When a bubble wall passes by the observer, he measures a space time
dependent charge potential for, say, fermionic hypercharge, which
induces transport phenomena and then local asymmetries in particle
numbers. The $Q'$, $(B-L)'$, and $BP'$ components of these asymmetries
remain unaffected by fast interactions, while the other
components are reprocessed as to obtain their
equilibrium values. In the case of $(B+L)'$ the
reprocessing is slow, so we have to use the rate equation
in (\ref{rate1}) to describe it. The generation of $(B+L)'_{SP}$ will go
on until either $\Gamma_{SP}$ goes to zero  or the local equilibrium value is
reached {\it i.e.} $(B+L)'_{SP} + (B+L)'_{TR} = (B+L)'_{EQ}$.
After the passage of the wall,  $(B+L)'_{EQ}$ and $(B+L)'_{TR}$ go
rapidly to zero since $\dot{\vartheta}$
vanishes, and so the final asymmetry is given by the $(B+L)'_{SP}$
generated until that time.

As we can see, the crucial question is now to calculate the induced
values for $Q'$, $(B-L)'$, $BP'$, and $(B+L)'_{TR}$. We will do that in the
next section by using linear response analysis \cite{LRT, Kapusta}.

\vspace{.5 cm}
{\Large\center\bf{3. Linear response analysis}}
\vspace{1. truecm}

In this paragraph we want to  discuss the effect on the thermal
averages of the term induced in the lagrangian when $\dot{\vartheta}$ is
active, which we assume to have the form
\beq
{\cal L} \rightarrow {\cal L} + \dot{\vartheta} J^0_{Y_F} ,
\label{cpot}
\eeq
with
\beq
J^0_{Y_F} = {\ts \sum _i}' y_{F}^i J^0_i
\label{hypercharge}
\eeq
where $\sum_i^{'}$ means that the sum extends on particles in
equilibrium with $\dot{\vartheta}$ only.
The standard procedure \cite{CKN} is to consider $\dot{\vartheta}$ as an
effective chemical potential, so that particle abundances at equilibrium
are given by
\beq
\rho_i = q_i \mu_Q + (b-l)_i \mu_{B-L} + bp_i \mu_{BP} + y^F_i
\dot{\vartheta},
\eeq
and then imposing
\beq
\langle Q'\rangle =\langle (B-L)'\rangle =\langle BP'\rangle =0
\eeq
so that any particle abundance can be expressed in terms of
$\dot{\vartheta}$ only. The point is that taking transport phenomena
into account, the above thermal averages are not zero, but depend on
$\dot{\vartheta}$ themselves. So we must first {\it calculate}
their values and then use them to determine the chemical potentials.
Our starting point is the generating functional
\beq
Z [J_{O}; \dot{\vartheta}] = \int_{P(A)BC} {\cal D}\phi
\exp\left\{i \int_{C} d\tau \int_V d^3\vec{x} \left[{\cal{L}} +
 \dot{\vartheta} J^0_{Y_F} + J_O O + `sources' \right] \right\}
\eeq
where ${\cal D}\phi$ is the integration measure, $O$ is the operator
of which we want to calculate the thermal average and $J_O$ the
corresponding source. $C$  is any path in the complex $\tau$ plane connecting
the point $\tau_{in}$ to $\tau_{out} = \tau_{in} - i \beta$
($\beta=1/T$) such that
the imaginary part of $\tau$ is never increasing on the path
\cite{TFT}. P(A)BC means that periodic (antiperiodic) boundary
conditions must be imposed on bosonic (fermionic) fields on the path.

The thermal average of the operator $O(\tau, \vec{x})$ in presence of
$\dot{\vartheta}$ is obtained in the usual way
\beq
\langle O(t_x, \vec{x})\rangle_{\dot{\vartheta}\neq 0} =
\frac{1}{i} \left.\frac{\delta Z [J_{O}; \dot{\vartheta}]}{\delta J_{O}
(t_x, \vec{x})}\right|_{J_O=0}.
\label{average}
\eeq
where all the field sources are set to zero.
Now we make a functional expansion of (\ref{average}) in
$\dot{\vartheta}$ and truncate it at the linear term,
\beq
\begin{array}{ccl}
\langle O(t_x, \vec{x})\rangle_{\dot{\vartheta}\neq 0} & = &\ds
\left.\frac{1}{i}\frac{\delta \:}{\delta J_O (t_x, \vec{x})} \right\{
Z [J_{O}; \dot{\vartheta}=0]\\
&&\\
&&\ds\left.\left. + \int_C d\tau' \int_V d^3\vec{y}
\:\dot{\vartheta}(\tau', \vec{y})\:\left.\frac{\delta Z [J_{O};
\dot{\vartheta}]}{\delta \dot{\vartheta} (\tau',
\vec{y})}\right|_{\dot{\vartheta}=0}
+\ldots \right\}\right|_{J_O=0}\\
&&\\
&=& \ds\langle O(t_x, \vec{x})\rangle_{\dot{\vartheta}=0}
+ i \int_C d\tau'\int_V d^3\vec{y} \: \dot{\vartheta}(\tau',
\vec{y}) \: \langle T_C
J^0_{Y_F}(\tau',\vec{y})\:O(t_x,
\vec{x})\rangle_{\dot{\vartheta}=J_O=0}
\end{array}
\label{linear}
\eeq
where $T_C$ is the ordering along the path C.

Different choices for the `time' contour $C$ lead to different
formulations of thermal field theory. One possibility is to take the
vertical line connecting $t_x$ to $t_x-i \beta$, so that
$\tau'-t_x$ is pure imaginary on any point of the path.
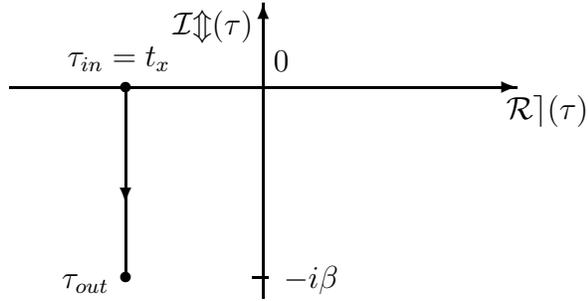
\begin{figure}
\begin{center}
\setlength{\unitlength}{0.4pt}%
\begin{picture}(500,280)(35,480)
\put(185,730){\makebox(0,0)[lb]{ ${\cal Im}(\tau)$}}
\put(280,695){\makebox(0,0)[lb]{ $0$}}
\put(500,650){\makebox(0,0)[lb]{ ${\cal Re}(\tau)$}}
\put( 85,700){\makebox(0,0)[lb]{ $\tau_{in}=t_x$}}
\put(80,490){\makebox(0,0)[lb]{ $\tau_{out}$}}
\put(290,490){\makebox(0,0)[lb]{ $-i\beta$}}
\thicklines
\put(150,680){\circle*{10}}
\put(150,680){\line( 0,-1){120}}
\put( 40,680){\vector( 1, 0){480}}
\put(150,680){\vector( 0,-1){110}}
\put(150,560){\line( 0,-1){ 60}}
\put(150,500){\circle*{10}}
\put(270,500){\line( 1, 0){ 15}}
\put(280,480){\vector( 0, 1){280}}
\end{picture}
\end{center}
\caption{\footnotesize  The path $C$ corresponding to imaginary time
formalism of thermal field theory.}
\end{figure}

This choice
corresponds to the imaginary time formalism (ITF) of thermal field
theory \cite{TFT}, and in this case we have to evaluate the euclidean
two point thermal Green's function $\langle T J^0_{Y_F} (z_E), O(0)
\rangle$ $(z_E^2 = -z_0^2 -|\vec{z}|^2)$. This can be done in
Matsubara formalism, where Feynman rules are straightforwardly
obtained \cite{Kapusta}. But, in this case, we would have to calculate
$\dot{\vartheta}$ for complex times, whereas we are interested in its
values at real times, during the passage of the wall. So, in order to
get a more direct physical interpretation of what we are calculating,
we must turn to real time formalism. This corresponds to  choose
the path in Fig. 2, and then letting $\tau_{in}$ going
to $-\infty$, and $t_F$ to $+ \infty$ \cite{TFT}.
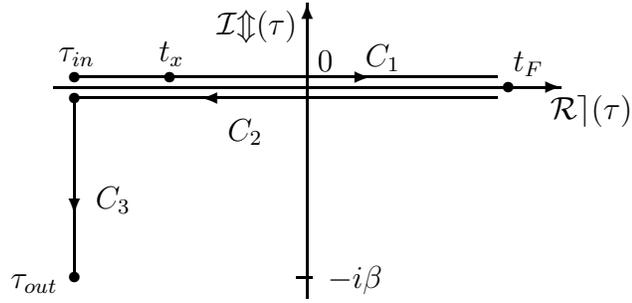
\begin{figure}
\begin{center}
\setlength{\unitlength}{0.4pt}%
\begin{picture}(500,280)(35,480)
\put( 70,565){\makebox(0,0)[lb]{ $C_3$}}
\put(195,630){\makebox(0,0)[lb]{ $C_2$}}
\put(185,730){\makebox(0,0)[lb]{ ${\cal Im}(\tau)$}}
\put(325,700){\makebox(0,0)[lb]{ $C_1$}}
\put(280,695){\makebox(0,0)[lb]{ $0$}}
\put(500,650){\makebox(0,0)[lb]{ ${\cal Re}(\tau)$}}
\put(465,695){\makebox(0,0)[lb]{ $t_{F}$}}
\put( 35,705){\makebox(0,0)[lb]{ $\tau_{in}$}}
\put(130,705){\makebox(0,0)[lb]{ $t_x$}}
\put(  -10,490){\makebox(0,0)[lb]{ $\tau_{out}$}}
\put(290,490){\makebox(0,0)[lb]{ $-i\beta$}}
\thicklines
\put( 60,690){\circle*{10}}
\put(150,690){\circle*{10}}
\put( 60,670){\circle*{10}}
\put( 60,500){\circle*{10}}
\put(470,680){\circle*{10}}
\put( 40,680){\vector( 1, 0){480}}
\put( 60,690){\vector( 1, 0){280}}
\put(340,690){\line( 1, 0){120}}
\put(460,670){\vector(-1, 0){280}}
\put(180,670){\line(-1, 0){120}}
\put( 60,670){\vector( 0,-1){110}}
\put( 60,560){\line( 0,-1){ 60}}
\put(270,500){\line( 1, 0){ 15}}
\put(280,480){\vector( 0, 1){280}}
\end{picture}
\end{center}
\caption{\footnotesize The path $C$ corresponding to real time
formulation of thermal field theory. $C_2$ lies infinitesimally beneath the
real axis.}
\end{figure}
Now, it is possible
to see that the contributions to the integral coming from $\tau'$ on
$C_3$ vanishes in the above limit \cite{TFT}, so we
are left with the contributions from $C_1$ and $C_2$ only. The $T_C$
ordering now allows us to rewrite the integral in (\ref{linear}) as
\beq
\begin{array}{c}
\ds i \int_{C_1\oplus C_2} d t_y\int_V d^3\vec{y} \:\dot{\vartheta}(t_y,
\vec{y})\:\langle T_C
J^0_{Y_F}(t_y,\vec{y})\:O(t_x,
\vec{x})\rangle_{\dot{\vartheta}=J_O=0}  \\
\\
\ds = i\int_{-\infty}^{t_x} d t_y \int_V d^3\vec{y} \:\dot{\vartheta}(t_y,
\vec{y})\:\langle \left[
J^0_{Y_F}(t_y,\vec{y}) , O(t_x, \vec{x})\right]
\rangle_{\dot{\vartheta}=J_O=0}   .
\end{array}
\eeq
Defining as usual the {\it retarded} Green's function
\beq
i D^R_{O,Y_F}(t_x, \vec{x}; t_y, \vec{y}) \equiv \langle \left[
O(t_x, \vec{x}), J^0_{Y_F}(t_y, \vec{y}) \right]\rangle
\Theta(t_x-t_y)
\eeq
where $\Theta(x)$ is the step function, we arrive at the result
\beq
\langle O(t_x, \vec{x})\rangle_{\dot{\vartheta}\neq 0} =
\langle O(t_x, \vec{x})\rangle_{\dot{\vartheta}= 0} +
\int_{-\infty}^{+\infty} d t_y \int_V d^3\vec{y}
\:\dot{\vartheta}(t_y,\vec{y})  D^R_{O,Y_F}(t_x, \vec{x}; t_y, \vec{y})
{}.
\label{response}
\eeq
The operators we are interested in are fermionic charge densities ($Q'$,
$(B-L)'$, $(B+L)'$, $BP'$) of the
form $Q_A' = \sum_i'q_i^A J^0_i$. Inserting it in (\ref{response}), and
using definition (\ref{hypercharge}) we
get
\beq
\langle Q_A'(t_x, \vec{x}) \rangle_{\dot{\vartheta}\neq 0} = {\ts
\sum_{ij}'}
q_i^A\:y^F_i \int_{-\infty}^{+\infty} dt_y \int_V d^3\vec{y}\:
\dot{\vartheta}(t_y, \vec{y})\: D^R_{i j}(t_x, \vec{x}; t_y, \vec{y})
\label{qui}
\eeq
where $D^R_{ij}$ is the current-current retarded Green's functions for fermion
$i$ and $j$ ($i$ and $j$ are flavour and color indices) and we used the
fact that  $\langle Q_A'(t_x, \vec{x}) \rangle_{\dot{\vartheta}=0} = 0$.
Note that the Green's function has to be evaluated for
$\dot{\vartheta}=0$, {\it i.e.} we must use the unperturbed lagrangian
with the charge potential turned off
and  all the chemical potentials equal to zero in the
partition function.

The problem of calculating the effect of the charge potential in
(\ref{cpot}) on the thermal averages for the particles in equilibrium
is then reduced  to the evaluation of the retarded Green's functions
which enter in (\ref{qui}). As we have discussed, the more natural framework
for this calculation is real time thermal field theory, in which the
physical sense of the various quantities is  evident. Anyway, we have
also seen that we can  calculate the euclidean Green's function in the
imaginary time formalism and then continue analytically to real times,
thus obtaining the $D^R_{ij}$'s (this relation was established for
the first time by Baym and Mermyn \cite{cont}).
In energy-momentum space the analytical continuation is accomplished
by the substitution
\beq
i \omega_n \rightarrow \omega + i \varepsilon\:\:\:\:\:\:\:
\varepsilon\rightarrow 0^+
\label{continuation}
\eeq
where $\omega_n=2 \pi n T $ are Matsubara frequencies and $\omega$ is
the real energy.

\vspace{.5 cm}
{\Large\center\bf{4. Solution of the rate equation}}
\vspace{1. truecm}

Since the rate of the sphaleronic transition is suppressed by
$\alpha_w^4$, the asymmetry in $(B+L)'$ generated by the sphalerons,
$(B+L)'_{SP}$,
is generally much smaller than both $(B+L)'_{EQ}$ and $(B+L)'_{TR}$ (we can
check it {\it a posteriori}), so we can approximate the rate equation in
(\ref{rate1}) by
\beq
\frac{d\:}{d t}{(B+L)'_{SP}} \simeq 0.92 \frac{\Gamma_{SP}}{T^3} \left[
(B+L)_{EQ}' - (B+L)_{TR}' \right].
\label{rate2}
\eeq
Using the equations (\ref{equilibrium}) and (\ref{qui})  we can
determine $(B+L)'_{LR}\equiv (B+L)'_{EQ}-(B+L)'_{TR}$ as
\beq
(B+L)'_{LR}(t_x, \vec{x}) = \langle J^0_{(B+L)'}(t_x, \vec{x})\rangle
= {\ts \sum_{ij}'} c_{ij}
 \int_{-\infty}^{+\infty} dt_y \int_V d^3\vec{y}
\:\dot{\vartheta}(t_y, \vec{y}) \: D^R_{i j}(t_x, \vec{x}; t_y, \vec{y})
\eeq
where
\[
c_{ij} \equiv y^F_j \left(
\frac{3}{80} q_i +\frac{7}{20} bp_i - \frac{19}{40} (b-l)_i -
(b+l)_i\right) .
\]

Integrating eq. (\ref{rate2}) in time from $-\infty$ to $+\infty$ we
get the final density of $(B+L)'$,
\beq
\begin{array}{ccl}
\Delta(B+L)'_{SP} &=&\ds \int_{-\infty}^{+\infty} d t_x
\frac{d\:}{d t_x}(B+L)'_{SP} (t_x, \vec{x})\\
&&\\
& = &\ds 0.92 \frac{2\pi}{T^3}
\int_{-\infty}^{+\infty} \:d\omega\: \widetilde{\Gamma}_{SP}(-\omega,
\vec{x})\:
\widetilde{(B+L)}'_{LR} (\omega, \vec{x})
\end{array}
\label{asim}
\eeq
where $\widetilde{\Gamma}_{SP}(-\omega, \vec{x})$ and
$\widetilde{(B+L)}'_{LR}
(\omega, \vec{x})$ are the Fourier transformed of
$\Gamma_{SP}(t_x, \vec{x})$ and  $(B+L)_{LR} (t_x, \vec{x})$ with
respect to time.

We recall that $\Gamma_{SP}$ is $k (\alpha_W T)^4$ in the unbroken
phase and decreases exponentially fast as the Higgs VEV is turned on.
In order to solve eq. (\ref{rate2}) analytically we approximate this
behaviour by a step function. Moreover, we will consider a plane
bubble wall moving along the $z$-axis with velocity $v_w$. So, our
expression for $\Gamma_{SP}$ will be
\beq
\Gamma_{SP}(t_x, \vec{x}) \simeq \Gamma \:\Theta\left( t_x - t_1 -
\frac{z_x}{v_w}\right)\: \Theta\left( t_2 - t_x + \frac{z_x}{v_w}\right)\:
\eeq
with $t_1\rightarrow -\infty$ and $\Gamma = k (\alpha_W T)^4$.
Of course, more sophisticated approximations for $\Gamma_{SP}$ may be
used, at the price of solving eq. (\ref{rate2}) numerically.

Analogously, we approximate $\dot{\vartheta}$ in such a way that it is
constant in a region of width $\Delta z$ inside the bubble wall, and
is zero outside,
\beq
\dot{\vartheta}(t_y, \vec{y}) = \theta \:\Theta(z_y - v_w t_y)\:
\Theta(v_w t_y- z_y +\Delta z)
\label{fase}
\eeq
where $\theta= v_w\:\Delta\vartheta/\Delta z$. So, if we are at  the
point $\vec{x}=0$, we observe an interaction of the form (\ref{cpot})
turned on from $t=-\Delta z/v_w$ to $t=0$, while the sphalerons are
active till $t=t_2$, as we have shown in Fig. 3.
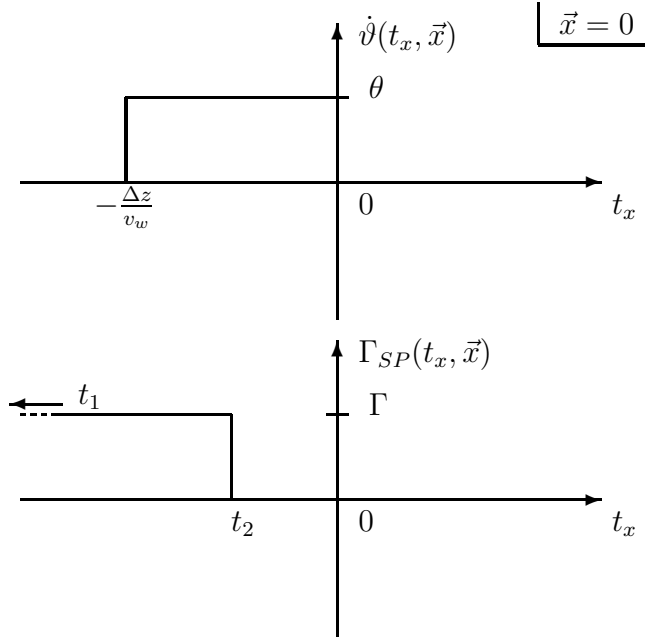
\begin{figure}
\begin{center}
\setlength{\unitlength}{0.4pt}%
\begin{picture}(570,580)(0,250)
\put(310,510){\makebox(0,0)[lb]{ $\Gamma_{SP}(t_x,\vec{x})$}}
\put(320,460){\makebox(0,0)[lb]{ $\Gamma$}}
\put(45,470){\makebox(0,0)[lb]{ $t_1$}}
\put(190,350){\makebox(0,0)[lb]{ $t_2$}}
\put(310,350){\makebox(0,0)[lb]{ $0$}}
\put(550,350){\makebox(0,0)[lb]{ $t_x$}}
\put(500,820){\makebox(0,0)[lb]{ $\vec{x}=0$}}
\put( 60,650){\makebox(0,0)[lb]{ $-\frac{\Delta z}{v_w}$}}
\put(310,650){\makebox(0,0)[lb]{ $0$}}
\put(550,650){\makebox(0,0)[lb]{ $t_x$}}
\put(310,810){\makebox(0,0)[lb]{ $\dot{\vartheta}(t_x,\vec{x})$}}
\put(320,760){\makebox(0,0)[lb]{ $\theta$}}
\thicklines
\put(490,810){\line(1,0){100}}
\put(490,810){\line(0,1){40}}
\put(300,760){\line(1,0){10}}
\put(100,760){\line(1,0){200}}
\put(100,760){\line(0,-1){80}}
\put( 0,680){\vector( 1, 0){550}}
\put(300,550){\vector( 0, 1){280}}
\put(300,250){\vector( 0, 1){280}}
\put(0,380){\vector( 1, 0){550}}
\put(200,460){\line( -1, 0){170}}
\put(290,460){\line( 1, 0){20}}
\put(200,460){\line( -1, 0){160}}
\put(35,460){\line( -1, 0){5}}
\put(25,460){\line( -1, 0){5}}
\put(15,460){\line( -1, 0){5}}
\put(5,460){\line( -1, 0){5}}
\put(40,470){\vector(-1,0){50}}
\put(200,460){\line( 0, -1){80}}
\end{picture}
\end{center}
\caption{\footnotesize Schematic representation of the behaviour of
$\dot{\vartheta}$ and of the rate of the sphaleronic transitions
at the point $\vec{x}=0$.}
\end{figure}

Putting all together we obtain
\beq
\begin{array}{cl}
\Delta(B+L)'_{SP}=& \ds 0.92 \frac{ (2 \pi)^3}{T^3} \Gamma \theta
{\ts \sum_{ij}'} c_{ij} \\
&\\
&\ds \int_{-\infty}^{+\infty} \:d\omega\: \frac{e^{-i\omega
t_2} - e^{-i \omega t_1}}{\omega} \frac{1 - e^{-i \omega \Delta z/v_w}}{\omega}
\:\widetilde{D}^R_{ij}(p_x=p_y=0, p_z=\frac{\omega}{v_w}, \omega) .
\label{asi}
\end{array}
\eeq
Note the peculiar relationship between $p_z$ and $\omega$ in the
argument of $\widetilde{D}^R_{ij}$, due to the spacetime dependence of
$\dot{\vartheta}(t_y, \vec{y})$, see (\ref{fase}).
A consideration of the general properties  of retarded Green's functions
\cite{Imagine} ensures that the imaginary part of $\widetilde{D}^R_{ij}$ is a
even function of $\omega$, while its imaginary part is odd. As a
consequence the integral in (\ref{asi}) will always give a real
result.

The lowest order contribution to $\widetilde{D}^R_{ij}$ comes from the
fermion loop in Fig. 4,
where the two crosses indicate the zero components of the fermion
current.
\begin{figure}
\begin{center}
\setlength{\unitlength}{1.pt}%
\begin{picture}(100,50)(0,0)
\put(-30,50){\makebox(0,0)[lb]{ $J^0_i$}}
\put(110,50){\makebox(0,0)[lb]{ $J^0_i$}}
\put(50,80){\makebox(0,0)[lb]{ $i$}}
\put(50,10){\makebox(0,0)[lb]{ $i$}}
\thicklines
\put(50,50){\oval(100,50)}
\put(-5,45){\line(1,1){10}}
\put(-5,55){\line(1,-1){10}}
\put(95,45){\line(1,1){10}}
\put(95,55){\line(1,-1){10}}
\put(50,75){\vector(1,0){0}}
\put(50,25){\vector(-1,0){0}}
\end{picture}
\end{center}
\caption{\footnotesize Lowest order contribution to $D^R_{ij}$}
\end{figure}
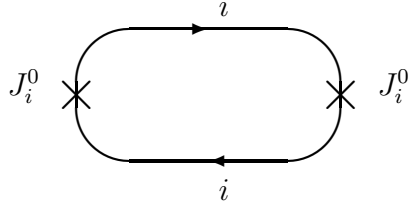
We may evaluate the corresponding euclidean two point Green's function in the
ITF and then continue analytically to real energies according to eq.
(\ref{continuation}). Moreover, since the relevant frequencies at
which $\widetilde{D}^R_{ij}$ must be  evaluated are such that $\omega \le
|\vec{p}| \le v_w/\Delta z \simeq T/10^2$ we take only the leading
terms in the high temperature expansion. These are given by \cite{Pis}
\beq
\Pi_l(p_0=2 n \pi T, \vec{p}) \:  \delta_{ij} =
 \frac{T^2}{3}\left[ 1 -\frac{i
p_0}{2 |\vec{p}|} \log\frac{i p_0 + |\vec{p}|}{i p_0 -
|\vec{p}|}\right]\: \delta_{ij}
\eeq
where  we have neglected fermion masses.
After continuing analytically to real energies and fixing the momenta
as in (\ref{asi}) we get the lowest order contribution to
$\widetilde{D}^R_{ij}$,
\beq
\widetilde{D}^{R\:0}_{ij} (p_x=p_y=0, p_z=\frac{\omega}{v_w}, \omega) =
\frac{T^2}{3} \left[1 - \frac{v_w}{2} \log \frac{1+v_w}{1-v_w} + i
\frac{\pi}{2} v_w  sign(\omega)\right] \delta_{ij} \: .
\label{pol}
\eeq
When
$v_w \rightarrow 1$ the above expression exhibits a collinear
divergence, due to the fact that the fermions in the loops are
massless in our approximation. This divergence disappears when plasma
masses for fermions are taken into account. However, for our purposes,
since $v_w \simeq
0.2$ \cite{thick}, the effects of plasma masses for fermions can be
neglected \cite{Lebedev}.

Due to the constraint $p_z=\omega/v_w$ the real part of the
correlation function in (\ref{pol})
does not depend on $\omega$, while the imaginary part depends  on
its sign only. This implies that the response induced on the plasma through
(\ref{pol}) has neither spatial nor temporal dispersion, {\it i.e.}
inserting (\ref{pol}) in (\ref{qui}) gives rise to an induced thermal
average for the charge $Q_A$ which in any space-time point is proportional to
the value of $\dot{\vartheta}$ in that point
\beq
\langle Q_A(t_x,\vec{x})\rangle^0 \propto \dot{\vartheta} (t_x, \vec{x})
,
\label{prop} .
\eeq
In particular also  $(B+L)_{TR}$ and $(B+L)_{EQ}$
receive a contribution of this form and disappear as soon as $\dot{\vartheta}$
is turned off. Inserting it into the rate
equation we obtain from (\ref{asi}) the contribution to the asymmetry
($t_1 \rightarrow - \infty$)
\beq
\Delta {(B+L)'_{SP}}^0 = 0.92 \frac{(2 \pi)^3}{T^3} \Gamma \theta
{\ts \sum_{ij}'} c_{ij} I^0(t_2) \delta_{ij}
\eeq
where
\beq
I^0(t_2) =  \frac{2 \pi}{3} T^2 \left(1 - \frac{v_w}{2}
\log\frac{1+v_w}{1-v_w}\right) \times \left\{
\begin{array}{lcc}
 0 &\:\:\:\:\:\:\:&  (t_2 < -\frac{\Delta z}{v_w})\\
&&\\
 \left(t_2 +\frac{\Delta z}{v_w}\right)
&\:\:\:\:\:\:\:& \left(-\frac{\Delta z}{v_w} < t_2 < 0\right)\\
&&\\
\frac{\Delta z}{v_w}
&\:\:\:\:\:\:\:& ( t_2 > 0)
\end{array}
\right.
\label{sol0}
\eeq
We recall that $t_2$ is the time at which sphaleron transitions are
turned off, while the charge potential induced by $\dot{\vartheta}$
is active for $-\Delta z/v_w < t < 0$. So the asymmetry calculated in
this approximation for $D^R_{ij}$ grows linearly with $t_2$ until
$t_2=0$ (for $t_2<-\Delta z/v_w$ the asymmetry is obviously zero since
there is no overlap between sphalerons and $\dot{\vartheta}$). The
result for $t_2>0$ is an artifact of our approximation
$(B+L)'_{SP}\ll(B+L)'_{LR}$, which is no more appropriate in this case.
Actually, from (\ref{prop}) we know that when  $\dot{\vartheta}$
goes to zero, as is the case for $t>0$, $(B+L)'_{EQ}$ and  $(B+L)'_{TR}$
 vanish too, and
the rate equation (\ref{rate1}) becomes
\beq
\frac{d\:}{d t} (B+L)'_{SP}
= - 0.92\frac{\Gamma_{SP}}{T^3} \left[ (B+L)_{SP} \right]
\:\:\:\:\:\:\:\:(t>0)
\label{rate3}
\eeq
so that the asymmetry produced before decreases exponentially from
$t=0$ to $t=t_2$ with rate $\Gamma$.
However, due to the smallness of $\Gamma$, and to the fact that
$t_2$ cannot be much larger than $O(\Delta z/v_w)$, we can safely
neglect this decreasing and take the result in   (\ref{sol0}).

Next, we consider the contribution to $D^R_{ij}$ due to gauge bosons
exchange. Since we are in the broken phase, and since  the perturbation
(\ref{cpot}) induced by  $\dot{\vartheta}$ is colorless, we will take
into account only photons, which contribute through the graph in Fig.
5.
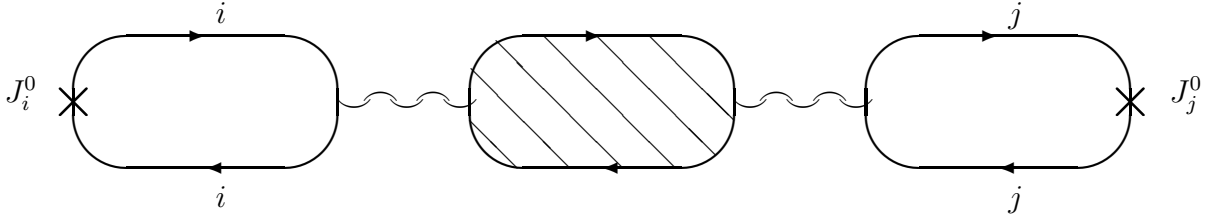
\begin{figure}
\begin{center}
\begin{picture}(80000,7000)(-8000,10000)
\put(-8000,16000){\makebox(0,0)[lb]{ $J^0_i$}}
\put(36100,16000){\makebox(0,0)[lb]{ $J^0_j$}}
\put(0,19000){\makebox(0,0)[lb]{ $i$}}
\put(0,12000){\makebox(0,0)[lb]{ $i$}}
\put(30000,19000){\makebox(0,0)[lb]{ $j$}}
\put(30000,12000){\makebox(0,0)[lb]{ $j$}}
\put(11800,13500){\line(-1,1){1800}}
\put(10050,17250){\line(1,-1){3750}}
\put(15800,13500){\line(-1,1){4800}}
\put(17800,13500){\line(-1,1){5000}}
\put(14800,18500){\line(1,-1){4500}}
\put(16800,18500){\line(1,-1){3200}}
\put(18800,18500){\line(1,-1){200}}
\thicklines
\drawline\photon[\E\FLIPPED](5000,16000)[5]
\drawline\photon[\E\FLIPPED](20000,16000)[5]
\put(0,16000){\oval(10000,5000)}
\put(15000,16000){\oval(10000,5000)}
\put(30000,16000){\oval(10000,5000)}
\put(-5500,15500){\line(1,1){1001}}
\put(-5500,16500){\line(1,-1){1001}}
\put(35500,15500){\line(-1,1){1001}}
\put(35500,16500){\line(-1,-1){1001}}
\put(0,18500){\vector(1,0){0}}
\put(0,13500){\vector(-1,0){0}}
\put(15000,18500){\vector(1,0){0}}
\put(15000,13500){\vector(-1,0){0}}
\put(30000,18500){\vector(1,0){0}}
\put(30000,13500){\vector(-1,0){0}}
\end{picture}
\end{center}
\caption{\footnotesize The contribution to $D^R_{ij}$ due to photon
exchange.}
\end{figure}
The blob in the middle represents the sum of all possible insertion of
fermion loops. In calculating the blob, we have to
include not only the fermions which enter in $Q'$, $(B-L)'$, and
$BP'$, {\it i.e.} the ones with fast flavour, or chirality,  changing
interactions, but we must instead take into account the contributions
of all the charged fermions of the theory. This is because QED
interactions are fast and then, for instance,  pair production is in
equilibrium
also for right handed light quarks.
The `QED' contribution of Fig. 5 gives
\beq
\widetilde{D}^{R,\:QED}_{ij} (\omega, \vec{p}) = e^2 q_i q_j \Pi^2_l(\omega,
\vec{p}) \frac{{\cal D}^{00}(\omega, \vec{p})}{1 - \sum_k (e q_k)^2
{\cal D}^{00} \Pi_l}
\label{QED}
\eeq
where $e q_i$ is the electric charge of the fermion $i$.
${\cal D}^{00}(\omega, \vec{p})$ is the tree level $(0, 0)$ component of the
photon propagator in the Coulomb gauge
\beq
{\cal D}^{\mu \nu} = \frac{-1}{p^2} P_T^{\mu \nu} -
\frac{1}{|\vec{p}|^2} u^\mu u^\nu \: ,
\eeq
where $p^2 = \omega^2-|\vec{p}|^2$, $u_\mu = (1,\:0,\:0,\:0)$
identifies the
plasma reference frame, while
\[
\begin{array}{c}
P_T^{00} = P_T^{0i} = P_T^{i0} = 0\\
\\
P_T^{ij} = \delta_{ij} - p^i p^j/|\vec{p}|^2
\end{array}
\]
As we have discussed, the sum in the denominator of (\ref{QED}) must
be extended over all the charged quarks and leptons.
Setting $p_z=\omega/v_w$ we get
\beq
\widetilde{D}^{R,\:QED}_{ij} (\omega, \vec{p})=-e^2 q_i q_j \:
v_w^2 \frac{\Pi_l^2(\omega,
p_z=\omega/v_w)}{\omega^2 + v_w^2 \sum_k (e q_k)^2 \Pi_l(\omega,
p_z= \omega/v_w)}.
\label{qed}
\eeq

Note that unlike the `direct' contribution (\ref{pol}) the `QED'  one is not
flavour (or color) diagonal, so that even particles which
do not enter into the expression for the charge potential (\ref{cpot}) get
a non zero thermal average depending on $\dot{\vartheta}$.
Moreover, this contribution has a  genuine $\omega$ dependence.
Two points retarded Green's function
are   analytic in the upper half of the complex $\omega$ plane.
$\widetilde{D}^{R,\:QED}_{ij} (\omega, p_z=\omega/v_w)$ may then have
poles of the
form $\overline{\omega} = \omega_p - i \gamma_p$, with $\gamma_p>0$.
In order to determine them we have to solve the following equations
\beq
\begin{array}{ccc}
\omega_p^2 - \gamma_p^2 &=& - C {\cal Re} \Pi_l(\overline{\omega})\\
&&\\
\omega_p & = &\ds  \frac{C {\cal Im} \Pi_l(\overline{\omega})}{2
\gamma_p}\\
&&
\end{array}
\label{polo}
\eeq
where $C=v_w^2 \sum_k (e q_k)^2$. Since the RHS of the first of eqs.
(\ref{polo})  is negative, and $v_w\simeq 0.2 <1$, we can approximate the
solutions by
\beq
\begin{array}{ccl}
\overline{\omega}_{1,\:2} &=&\pm \omega_p + i \gamma_p\\
&&\\
\omega_p &=& \ds \frac{\pi}{4} v_w \gamma_p\\
&&\\
\gamma_p &\simeq &\ds \left[ C {\cal Re}
\Pi_l(\overline{\omega})\right]^{1/2} \simeq  v_w \frac{e T}{\sqrt{3}}
\left(\sum_k (q_k)^2\right)^{1/2}\\
&&
\end{array}
\label{poles}
\eeq
where $\gamma_p >0$ as it should be.

Inserting (\ref{qed}) in eq. (\ref{asi}) we obtain the `QED'
contribution to the asymmetry
\beq
\Delta {(B+L)'_{SP}}^{QED} \simeq 0.92  \frac{(2 \pi)^3}{T^3} \Gamma \theta
{\ts \sum_{ij}'} c_{ij} I^{QED} (t_2)
\eeq
where, now,
\beq
\begin{array}{cl}
I^{QED} (t_2) =&\ds - \frac{2 \pi}{3} T^2 \left(1 - \frac{v_w}{2}
\log\frac{1+v_w}{1-v_w}\right)\frac{q_i q_j}{\sum_k (q_k)^2}
\\
&\\
\times &\left\{
\begin{array}{lcc}
 0 &\:\:\:&  (t_2 < -\frac{\Delta z}{v_w})\\
&&\\
\ds  \left[t_2 +\frac{\Delta z}{v_w} + \cos\omega_p\left(t_2 +
\frac{\Delta_z}{v_w}\right) \frac{e^{-\gamma_p \left(t_2 +
\Delta_z/v_w\right)}}{\gamma_p}\right]
&\:\:\:& \left(-\frac{\Delta z}{v_w} < t_2 < 0\right)\\
&&\\
\ds  \left[\frac{\Delta z}{v_w} -
\frac{e^{-\gamma_p t_2}}{\gamma_p}\left(\cos\omega_p t_2
- e^{-\gamma_p \Delta z/v_w}
\cos\omega_p\left(t_2+ \frac{\Delta z}{v_w}\right)\right)\right]
&\:\:\:& ( t_2 > 0) ,
\end{array}\right.\\
&
\end{array}
\label{solqed}
\eeq
The same considerations about the case $t_2>0$ made after eq.
(\ref{sol0}) apply also here. We can notice that photon exchange gives
two different types of  contributions. The first one has the same
behaviour of that in (\ref{sol0}), {\it i.e.} a linearly increasing
asymmetry from $t_2=-\Delta z/v_w$ to $t_2=0$. On the other hand, the second
term exhibits a well known feature of plasma physics, namely, plasma
damped oscillations induced by an external perturbation
\cite{Kapusta}. The damping rate here is given by $\gamma_p$.
In the case $-\Delta z/v_w < t_2<0$, we see that the oscillating term
dominates over the linear one only for $t_2 \rightarrow - \Delta z/v_w$,
and is rapidly damped as $t_2\rightarrow 0$, since $\exp(-\gamma_p
\Delta z/v_w)\simeq$ $\exp (-T\Delta z)\simeq$ $\exp -(40)$.
When $t_2>0$ the amplitude of the oscillation is always suppressed at
least by a factor $10^{-1}\div 10^{-2}$ with respect to the linear
one, and then we can conclude that the effect of the oscillating term is
negligible unless $t_2$ is very near to $-\Delta z /v_w$.

An interesting feature of our results (\ref{sol0}) and (\ref{solqed})
can be appreciated if we calculate, by means of eq.
(\ref{qui}), the electric charge $Q'$  induced by the phase
$\dot{\vartheta}$, taking into account both the direct contribution (\ref{pol})
and the `QED' one (\ref{QED}) to $\widetilde{D}^R_{ij}$.
It is easy  to see that it is given by
\beq
\begin{array}{cl}
\langle Q' \rangle \propto & \left[\sum_i^{'} y_i q_i \left(\sum_k q_k^2 -
\sum_k^{'}  q_k^2\right) \times ``linear \:\:contribution"\right]\\
&\\
&+ ``damped\:\: contribution",
\end{array}
\label{carica}
\eeq
then, when {\it every} fermion is in equilibrium, the linear
contribution to the thermal average of $Q'$ vanishes, and we are left
with the damped one. The reason is that in this case $Q'$ coincides
with the total fermion electric charge, and this is perfectly screened
as in the usual QED plasma. Since in the real situation  not all the
fermions are in equilibrium, the linear contribution to (\ref{carica})
does not cancel. Anyway, this considerations are valid for electric charge
only, while the other interesting charges, $(B+L)'$, $(B-L)'$, and, $BP'$,
would have non zero linear contributions even if all the fermions were
in equilibrium.

Putting all together, and neglecting the damped contribution, we
find the final asymmetry in $(B+L)'$
\beq
\Delta(B+L)'_{SP}= - 0.92 \frac{37}{240}(2\pi)^4 \frac{\Gamma
\theta}{T} \left(1-\frac{v_w}{2}
\log\frac{1+v_w}{1-v_w}\right) \left(t_2 + \frac{\Delta
z}{v_w}\right),
\label{asu}
\eeq
where we have assumed that the sphalerons turn off when the phase is
still active ($-\Delta z/v_w <$  $t_2 <0$).
Recalling that $\theta = $ $\Delta\vartheta v_w/\Delta z$, and assuming
that sphalerons cease to be active after a time interval $t_2 +\Delta
z/v_w =$ $f\Delta z/v_w$ from the turning on of $\dot{\vartheta}$ we get
\beq
\Delta(B+L)'_{SP} \simeq -2.3\cdot 10^{2} k T^3 \alpha_w^4 \Delta
\vartheta f
\label{bau}
\eeq
where we have taken the reasonable value $v_w\simeq 0.2$
\cite{thick}. The above value is enhanced by nearly three orders of magnitude
with respect to the original estimates by CKN \cite{CKN} where transport
phenomena were not taken into account.

The predicted baryon asymmetry of the Universe then comes out to be
\beq
\frac{\rho_B}{S} \simeq -10^{-6} k \Delta \vartheta f.
\eeq
$k$ is estimated in the range $0.1 \div 1$ from numerical simulations
\cite{k}, while the value of $f$ is still an open question.
Following Dine and Thomas \cite{dinethomas} we chose
\beq
f\simeq\frac{\alpha_w}{g}\simeq 5\cdot 10^{-2}.
\eeq
The observed baryon asymmetry, $\rho_B/S = (4\div7)\cdot 10^{-11}$, can then
be reproduced for $\Delta\vartheta\simeq 10^{-2} \div 10^{-3}$,
values which can be
obtained either by explicit CP violation or by spontaneous CP
violation at finite temperature \cite{scpv} without entering in
conflict with the experimental bounds on the electric dipole moment of
the neutron.

The result (\ref{asu}) was obtained considering a charge potential of the form
(\ref{cpot}) where the sum extends on all the left handed fermions
plus the right handed quark. Considering the more physical situation in
which only the top quarks (left and right handed) feel the effect of
$\dot{\vartheta}$ the coefficient $37/240$ in (\ref{asu}) should be changed
into $9/32$, thus leading to an enhancement of a factor $1.8$.

\vspace{.5 cm}
{\Large\center\bf{5. The effect of QCD sphalerons.}}
\vspace{1. truecm}

It is well known that the axial vector current of QCD has a triangle anomaly,
therefore one can expect axial charge violation due to topological transitions
analogous to the sphaleronic transitions of the electroweak theory.
The rate of these processes at high temperature may be estimated as
\cite{QCD, GianShap}
\beq
\Gamma_{strong} = \frac{8}{3} \left(\frac{\alpha_s}{\alpha_W}\right)^4
\Gamma_{SP} = \frac{8}{3} k  (\alpha_s T)^4
\eeq
where $\alpha_s$ is the strong fine structure constant, leading to  a
characteristic  time of order
\beq
\tau_{strong} =\frac{1}{192 k \alpha_s^4 T}.
\eeq
Since $k\simeq 0.1\div 1$ \cite{k}, we see that $\tau_{strong}$ is
comparable to the time of passage of the bubble wall, and might
also be smaller.

Recently, Giudice and Shaposhnikov have analyzed the effect of these `QCD
sphalerons' on the adiabatic baryogenesis scenario.
They showed that, as long as these transitions are in equilibrium and
fermion masses are neglected, no baryon asymmetry can be generated.
Thus, the final result will be suppressed by a factor $\sim (m_{top}(T)/\pi
T)^2$. In this paragraph we will reconsider the issue taking transport
phenomena into account.

The effect of QCD sphalerons may be represented by the operator
\beq
\Pi_{i=i}^{3}(u_L\: u_R^\dagger\: d_L\: d_R^\dagger)_i
\eeq
where $i$ is the generation index. Assuming that these processes
are in equilibrium, we get the following chemical potentials equation
\beq
\sum_{i=1}^3 (\mu_{u_L^i} - \mu_{u_R^i} + \mu_{d_L^i} - \mu_{d_R^i}) = 0
\label{QCDs}.
\eeq
Eq. (\ref{QCDs}) contains the chemical potentials for {\it all} the
quarks, and imposes that the total {\it right-handed} baryon
number is equal to the total {\it left-handed} one.
Using eqs. (\ref{chem1})
and (\ref{chem2}) we can rewrite it as
\beq
4 \mu_{u_L} + \mu_{t_L} - \mu_{b_R} - 2 \mu_{d_R} - 2 \mu_{u_R} - 3
\mu_{W^+} = 0,
\label{qs}
\eeq
where $\mu_{u_{L,R}}\equiv 1/2\sum_{i=1}^2 \mu_{u_{L,R}^i}$, and
$\mu_{d_{R}}\equiv 1/2\sum_{i=1}^2 \mu_{d_{R}^i}$. One of the three new
chemical potentials, $\mu_{b_R}$, $\mu_{d_R}$, and $\mu_{u_R}$, can be
eliminated using eq. (\ref{qs}), while the remaining two correspond to
two more conserved charges that must be taken into account besides
$Q'$, $BP'$, and $(B-L)'$ (now the primes mean that the summation has
to be performed on right handed quarks too, but not on right handed
leptons). We can choose
\beqra
X &\equiv & \sum_{i=1}^3 d_R^i - \frac{3}{2}\sum_{i=1}^2 u_R^i,\\
Y &\equiv & b_L + t_L + t_R + \frac{1}{2} \sum_{i=1}^2 u_R^i
- \sum_{j=1}^3 (e_L^j + \nu_L^j),
\eeqra
corresponding respectively to $A_3$ and $A_2$ in the notation of ref.
\cite{GianShap}.
Following the usual procedure, we can now express the abundance of any
particle number at equilibrium as a linear combination of $Q'$,
$(B-L)'$, $BP'$, $X$, and $Y$. For $(B+L)'$ we obtain the result
\beq
(B+L)_{EQ}'=- \frac{1}{5} (B-L)',
\label{aa}
\eeq
to be compared to eq. (\ref{equilibrium}), which we obtained considering QCD
sphalerons out of equilibrium. Thus, the equilibrium value for
$(B+L)'$ depends only on the the density of $(B-L)'$, in agreement with
what was obtained in ref. \cite{GianShap}\footnote{Incidentally, note that
this is not
a general property due to the insertion of QCD sphalerons into the set of
processes in equilibrium, but is due to the fact that only top Yukawa
interactions are fast. If, for instance, bottom quark Yukawa interactions
were also fast, then we would find that $(B+L)'_{EQ}$ is not simply
proportional to
$(B-L)'$, but depends also on the values of the other charges in equilibrium.}.
If transport phenomena
were not present, as it was assumed in ref. \cite{GianShap}, we could
set $(B-L)'$ to zero and then conclude that QCD sphalerons allow no
non vanishing $(B+L)'$ density, at equilibrium and in the massless
approximation.
On the other hand, including transport effects, we can easily
calculate the $(B-L)'$ density induced by $\dot{\vartheta}$ using eq.
(\ref{qui}), and then, through (\ref{aa}), the equilibrium value
$(B+L)'_{EQ}$, which, unlike in ref. \cite{GianShap}, comes out to be non
vanishing inside the bubble wall. However this is not sufficient to
conclude that we will
have a non zero final baryon asymmetry when the bubble wall has passed by.
As we discussed in Sect. 2., the
generation of baryons inside the bubble walls is described by eq.
(\ref{rate1}), with the initial condition $(B+L)'_{SP}=0$.
Then we must calculate $(B+L)'_{TR}$, {\it i.e.} the contribution to
$(B+L)'$ due to transport. If all the particles in equilibrium participated to
the charge potential then we would find
\beq
(B+L)'_{TR}(t_x, \vec{x}) = (B+L)'_{EQ}(t_x, \vec{x})
\label{azzo}
\eeq
so that the system would always be on the minimum of the free energy inside the
bubble  wall and there would be no bias of the (electroweak) sphaleronic
transitions.
As a consequence, $(B+L)_{SP}$ would remain
zero and no asymmetry would survive after the passage of the wall up to fermion
mass effects, in agreement with what was find in ref. \cite{GianShap}.

On the other hand, including only left and right handed top quarks into the
charge potential, eq. (\ref{azzo}) is no more satisfied and a non zero result
for the final baryon asymmetry  is recovered. In this case, the factor
$37/240$ in eq. (\ref{asu}) should be changed to $25/72$.

\vspace{.5 cm}
{\Large\center\bf{6. Conclusions}}
\vspace{1. truecm}

In this paper we have analyzed the impact of transport phenomena on the so
called `spontaneous baryogenesis' mechanism of Cohen, Kaplan, and Nelson.
We have assumed that inside the walls of the bubbles nucleated during the
electroweak phase transition a space time dependent `charge potential'
for (partial) fermionic hypercharge is generated. We stress again that no
discussion about the origin of the charge potential has been given; in
particular, since in the limit in which all the Yukawa couplings go to
zero there is no communication between the Higgs and the fermion
sectors,  we expect that in this limit also the charge potential
should  go to zero. In the traditional approach of CKN there is no
trace of this behaviour, and moreover we have shown that the results change
in a sensible way according to the precise form of the charge potential which
is considered. We reserve a discussion on this subject
for a forthcoming publication. Our main interest here was
to set a scheme for calculations in the
the adiabatic scenario  in the case in which such a charge potential
is present, in order  to determine the variations in the thermal
averages induced by transport effects and the production of baryon
number by sphalerons inside the bubble walls.

The main physical point of the paper is that the system should be
regarded as a collection of subsystems in local
equilibrium, in which the thermal averages for the conserved charges
are not zero but are driven to non vanishing values by transport
phenomena. In particular, the local equilibrium configurations will correspond
to non zero values for $(B+L)'$.
We have determined the local equilibrium configuration by means of a
chemical potential analysis, calculating the values of the thermal
averages for the conserved charges by using linear response theory.
We have considered only the dominant contribution to these averages,
in particular, we have neglected any fermion mass effect and also the
coupling of the Higgs field to the Chern-Simons number.

We find that, in contrast with previous claims \cite{JPT}, the presence of
transport phenomena does not prevent baryon number generation inside
the bubble walls. The main source of disagreement with JPT is the
following. In their paper, JPT consider the rate equation in
the form
\beq
\dot{B} = -\frac{\Gamma_{SP}}{2T} \:(3 \mu_{t_L} + 3 \mu_{b_L} +
\mu_{\tau_L} + \mu{\nu_\tau} ),
\eeq
where the term on the right hand side has been obtained by considering
the variation of the free energy of the system due to a
`sphaleron-like'  transition involving only the third generation, {\it
i.e.} due to the processes $t_L t_L b_L \tau_L \leftrightarrow 0$
and  $t_L b_L b_L \nu_{\tau} \leftrightarrow 0$. Then these authors
impose that the chemical potential of any particle is proportional to the
value of its  hypercharge, and so they find that the right hand side
vanishes as a consequence of the conservation of hypercharge (and of
fermion hypercharge) in any sphaleronic transition.  The point is
that, assuming local equilibrium of the fast interactions,
the chemical potentials of the single particle species {\it are not}
proportional to their hypercharge. In fact, since the single particle
numbers are not conserved quantities of the system, their abundances
are reprocessed by fast interactions as to obtain their local equilibrium
values. On the
other hand, imposing that the particle chemical potentials are
proportional to hypercharge, would be equivalent to freeze out any
interaction inside the bubble wall, both the fast and the slow ones,
leaving transport phenomena as the only relevant process.

Also, if a charge potential for fermionic or total hypercharge is
present, transport phenomena allow the generation of the
baryon asymmetry even in the limit in which the Higgs VEV's go to
zero. The reason is again that the thermal averages for $Q'$, $(B-L)'$ and
$BP'$ are non vanishing and then a $(B+L)$ asymmetry can be generated
even if the  electroweak symmetry is unbroken. This is strictly
analogous to the well known result of the survival of a $B+L$
asymmetry when a $B-L$ density, eventually of GUT origin, is present
\cite{HarveyDreiner}. Of course, in the limit in which the VEV's go to
zero, also the charge potential should go to zero, since no complex phase can
emerge from the Higgs sector in this case. Then, also this suppression,
like the one due to vanishing $h_{top}$,  should be made evident by an accurate
discussion on the origin of the charge potential. As a
consequence, our results for the baryon asymmetry, eq. (\ref{bau})
should be probably  multiplied by a further suppression factor roughly
of order $(m_{top}(T)/T)^2$.

\vspace{1. cm}
\centerline{\bf Acknowledgments}
One of us (M.P.) is very grateful to the DESY theory group for the
kind hospitality during the last stage of this work.

\end{document}